\documentclass[pre,superscriptaddress,twocolumn,showpacs,amsmath,amssymb,floatfix,amstex,footinbib]{revtex4-1}

\usepackage{units}
\usepackage{psfrag}
\usepackage{mathtools} 

\usepackage{graphicx}

\newcommand{\rmd}{\mathrm{d}}

\newcommand{\dn}{\Delta n^2}

\newcommand\energy{E}
\newcommand\norm{\mathcal{N}}

\newcommand{\mysection}[1]{\section{#1}}
\newcommand{\mysubsection}[1]{\subsection*{#1}}
\newcommand{\new}[1]{#1}

\begin{document}

\title{Scaling properties of energy spreading in nonlinear Hamiltonian two-dimensional lattices}

\author{Mario Mulansky} 
\affiliation{Institute of Physics and Astronomy, University of Potsdam, 
  Karl-Liebknecht-Str 24, D-14476, Potsdam-Golm, Germany}
\affiliation{Institut Henri Poincar\'e, 11 rue Pierre et Marie Curie, 75005 Paris}
\author{Arkady Pikovsky} 
\affiliation{Institute of Physics and Astronomy, University of Potsdam, 
  Karl-Liebknecht-Str 24, D-14476, Potsdam-Golm, Germany}

\date{\today}

\begin{abstract}
\noindent In nonlinear disordered Hamiltonian lattices, where there are no propagating phonons, the spreading of energy is of subdiffusive nature. Recently, the universality class of the
subdiffusive spreading according to the nonlinear diffusion equation (NDE) has been suggested and checked for one-dimensional lattices.
Here, we apply this approach to two-dimensional strongly nonlinear lattices and  find a nice agreement of the scaling predicted from the NDE with the spreading results from extensive numerical studies. Moreover, we show 
that the scaling works also for regular lattices with strongly nonlinear coupling, for which the scaling exponent is estimated analytically.
This, for the first time, shows that the process of chaotic diffusion in such lattices does not require disorder.
\end{abstract}


\maketitle

\mysection{Introduction}

Nonlinearity may have a nontrivial effect on energy spreading in extended systems. In regular, homogeneous in space media and lattices, nonlinearity can partially block or reduce the linear spreading mediated by waves (phonons), due to creation of localized structures (solitons, breathers). Contrary to this, in systems where there are no propagating linear waves, nonlinearity is the only mechanism responsible for the spreading of initially localized wave packets. This situation will be addressed in this work. One of the possibilities to suppress the linear spreading is to introduce disorder in the system. Then, in one and two dimensions, all linear eigenmodes become localized and their spectrum becomes purely discrete, this effect is known as Anderson localization (see~\cite{Abrahams-10} and~\cite{Lye-05,Schwartz-07,Lahini-08} for recent experimental observations). In this case nonlinearity leads to a weak subdiffusive spreading of the wave packets, as has been demonstrated in one~\cite{%
Shepelyansky-93,Molina-98,%
Pikovsky-Shepelyansky-08,%
Garcia-Mata-Shepelyansky-09,Garcia-Mata-Shepelyansky-09a,%
Flach-Krimer-Skokos-09,Laptyeva-10,%
Mulansky-Ahnert-Pikovsky-Shepelyansky-09,%
Fishman-Krivolapov-Soffer-12} and two~\cite{Garcia-Mata-Shepelyansky-09,Laptyeva-Bodyfelt-Flach-12} dimensions. It should be noted that these results are mainly based on numerical experiments, while purely theoretical attempts (cf.~\cite{Bourgain-Wang-07,Basko-11,Fishman-Krivolapov-Soffer-09}) have not been fully 
successful in explaining numerical observations. 
Also, by studying chaos properties~\cite{Pikovsky-Fishman-11} and the possible existence of KAM tori~\cite{Johansson-Kopidakis-Aubry-10}, some reasoning for a slowing down of spreading has been proposed.

Another situation where linear waves are absent are so-called strongly nonlinear lattices~\cite{Ahnert-Pikovsky-09,Mulansky-Ahnert-Pikovsky-11,Roy-Pikovsky-11}. These lattices consist of linear or non-linear oscillators which are coupled to nearest neighbors by nonlinear forces. The propagating modes here can be  nonlinear waves only, typically these waves are
compactons~\cite{Rosenau-Hyman-93}. An example of such a lattice is the well-known toy ``Newton's cradle''~\cite{Sen-Hong-Bang-Avalos-Doney-08}. Again, disorder in such a lattice blocks the compactons (if they exist) and the observed spreading is not a traveling wave phenomenon, but a slow subdiffusive process.

All the numerical evidence in the cited literature indicates for a subdiffusive spreading of wave packets due to weak chaos. A natural question is whether this phenomenon
can be described phenomenologically as a certain ``universality class'', similar to successful
statistical approaches to problems like percolation and roughening of interfaces. In recent papers~\cite{Mulansky-Ahnert-Pikovsky-11,Mulansky-Pikovsky-12} such a phenomenological description based on the properties of the Nonlinear Diffusion Equation (NDE) has been proposed and tested for one-dimensional lattices.
Here, we extend this theory to two dimensions, and compare the scaling following from the NDE with numerical results from an extensive study on 2D strongly nonlinear lattices.
We emphasize that within this framework we also can investigate spreading in regular lattices, where all oscillators are in resonance. This is 
a new, theoretically important case helping to promote the understanding of nonlinear spreading, as here also a theoretical prediction on the spreading exponent is possible.

We start in section~\ref{sec:2dsnl} by describing the two-dimensional Hamiltonian lattices.
Then, in section~\ref{2dnde}, we introduce the two-dimensional NDE and 
deduce the scaling and spreading properties from its self-similar solution.
This is followed by section~\ref{sec:nr} with our main results, the comparison of these predictions with the extensive numerical simulations including theoretical predictions of the spreading exponents when possible.

\mysection{2D Strongly Nonlinear Lattices}
\label{sec:2dsnl}
The model considered here is a straightforward generalization of the system considered in~\cite{Mulansky-Ahnert-Pikovsky-11,Mulansky-Pikovsky-12} to two dimensions; it consists of oscillators with power-law local and interaction potentials  described by the following Hamiltonian:
\begin{equation}
\begin{aligned}
 H &= \sum_{i,k} E_{i,k}\\
  E_{i,k}&= \frac{p_{i,k}^2}{2} + \frac{W\omega_{i,k}^2}{\kappa} |q_{i,k}|^\kappa+\\ &+ \frac{\beta}{2\lambda} (|q_{i+1,k} - q_{i,k}|^\lambda+|q_{i-1,k} - q_{i,k}|^\lambda \\
 & \qquad + |q_{i,k+1} - q_{i,k}|^\lambda+ |q_{i,k-1} - q_{i,k}|^\lambda)\;.
\end{aligned}
\label{eqn:general_hamiltonian}
\end{equation}
Here $q_{i,k}$ and $p_{i,k}$ are the positions and momenta of the oscillator at lattice site $i,k$ of a quadratic lattice with nearest neighbor coupling.
Parameters
$\kappa$ and $\lambda$ represent the power of the local and the coupling potentials; $\omega_{i,k}$ is a parameter of the local potential and we study this model with a random local potential (randomly iid.\ $\omega_{i,k} \in [0,1]$) as 
well as in the regular case where $\omega_{i,k} = 1$. Parameters
$W$ and $\beta$ describe the local and the coupling strength in this model.

We have a freedom of rescaling the Hamiltonian and the time to get rid of some parameters (this is done similarly to the one-dimensional case~\cite{Mulansky-Ahnert-Pikovsky-11,Mulansky-Pikovsky-12}).
In the case of different nonlinear powers $\kappa \neq \lambda$,  
parameters $W$ and $\beta$ can be set to $W=\beta=1$ by rescaling $q$, $p$ and $t$, and the only remaining parameter is the total energy $\energy$ in the system.

For a homogeneous nonlinearity with equal powers $\kappa=\lambda$, 
by a proper rescaling, the energy and the local nonlinear strength can be set to $\energy=W=1$ and $\beta$ is the only remaining parameter in this case, describing the relation of the coupling and the local potential.
This possibility to scale the total energy to unity induces the following scaling relation between the energy and the time~\cite{Mulansky-Ahnert-Pikovsky-11,Mulansky-Pikovsky-12}: 
\begin{equation} \label{eqn:energy_time_scaling}
 t \sim \energy^{1/\kappa-1/2} .
\end{equation}
We will use this result later for comparison with the scaling predictions of NDE.

\begin{figure}[t]
 \centering
  \includegraphics[width=0.48\textwidth]{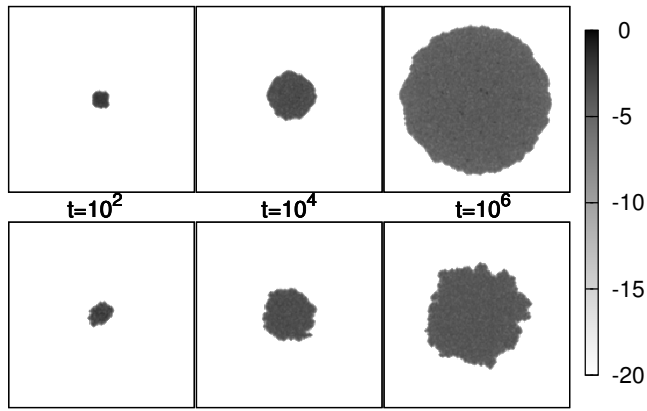}
 \caption{Instantaneous logarithmic local energy density $\log_{10} w_{i,k}$ for $\kappa=2$, $\lambda=4$ at times $10^2$, $10^4$ and $10^6$ (left to right panels). The upper row shows results for a regular lattice ($\omega_{i,k}=1$) with energy $\energy=1$ ($W=\beta=1$ from variable transformations). The lower row is from simulations of a disordered lattice ($\omega_{i,k} \in [0,1]$) with energy $\energy=10$. The total size of the squares is 160x160 lattice sites.
 \label{fig:2_4_wf_2d}
 }
\end{figure}

Starting from initially localized excitations in such a two-dimensional lattice, we find sharply localized spreading states.
This observation is illustrated in Fig.~\ref{fig:2_4_wf_2d} where we show the time evolution of a random initial excitation of 5x5 sites for a lattice with $\kappa=2$, $\lambda=4$ and random local potential ($\omega_{i,k} \in [0,1]$) as well as the regular case ($\omega_{i,k}=1$).
Our fundamental observable is the distribution of energy density $w_{i,k} = E_{i,k}/\energy$ where $E_{i,k}$ is the local energy at site ${i,k}$ defined in~(\ref{eqn:general_hamiltonian}).
One observes in Fig.~\ref{fig:2_4_wf_2d} that the extent of the excitation clearly increases with time and has very sharp tails at the boundary (note the logarithmic scale of the grey coding).
The states spread roughly circular, although with disorder (lower row in Fig.~\ref{fig:2_4_wf_2d}) the boundary is more rough than in the regular case.
We note that, opposed 
to the 1D case ~\cite{Mulansky-Pikovsky-12,Ahnert-Pikovsky-09},
no compactons or quasi-compactons that propagate through the lattice are observed in the 2D case, even without disorder (upper row in Fig.~\ref{fig:2_4_wf_2d}). This allows us to apply the same approach to both regular and disordered lattices.

To quantify the spreading of energy density shown in Fig.~\ref{fig:2_4_wf_2d} we introduce the 2D second moment calculated as:
\begin{equation}
 \dn 
  = \sum_{i,k} \left( (m_x - i)^2 + (m_y - k)^2 \right) w_{i,k}.
\end{equation}
with $m_{x} = \sum_{i,k} i\, w_{i,k}$ and $m_{y} = \sum_{i,k} k\, w_{i,k}$ being the center of the distribution.
$\dn$ is a measure for the excitation area in terms of the number of excited sites. In order to have a characterization of the uniformity of the wave field,
we need to calculate other quantities that are more or less sensitive to peaks in the distribution. Following~\cite{Varga-Pipek-03} we calculated the R\'enyi
entropies of the distribution $S=-\sum_{i,k} w_{i,k}\ln w_{i,k}$ and $S_2 = -\ln \sum_{i,k} 1/w_{i,k}^2$, (the latter is directly related to the participation number), and combined them into the structural entropy defined as:
\begin{equation}
 S_\text{str} = S - S_2\;.
\end{equation}
If the structural entropy is constant in the course of evolution, then the relative strength of the peaks in the distribution of energy does not change; a growing structural entropy means that peaks become relatively stronger.

\mysection{2D NDE}
\label{2dnde}

The Nonlinear Diffusion Equation (NDE) was found to give reasonable predictions on the spreading in one-dimensional strongly nonlinear lattices~\cite{Mulansky-Ahnert-Pikovsky-11,Mulansky-Pikovsky-12}.
The general picture one has in mind when assuming the NDE to describe spreading states is that the motion of already excited, active, lattice sites is chaotic.
This can be viewed as an intrinsic stochasticity of the motion which allows for new lattice sites to get excited by stochastic driving from the active neighbors.
The chaos, and therefore also the excitation efficiency, gets weaker in the course of spreading, because the energy density decreases.
This effect is accounted for by introducing a density dependent diffusion coefficient, in terms of a simple power-law of the energy density ~$D(w)\sim w^a$, inspired from the power-law nonlinearities of the model.
This is the physical argument for employing the NDE in nonlinear systems that exhibit some sort of chaotic diffusion.
However, a rigorous derivation is yet to be found and such a calculation is not aimed at in this work.

Below, we want to present an even wider applicability of the NDE to 2D systems and fully resonant oscillators that has not been investigated before.
It should be noted here already that this understanding of chaotic diffusion as the mechanism of the spreading process does not rely on the presence of disorder.
The disorder is only essential for blocking non-linear waves in such system with purely nonlinear coupling.
However, in 2D no such waves exist, c.f.~Fig.~\ref{fig:2_4_wf_2d}, and one can study chaotic diffusion in regular lattices as well, as is shown later.

But first, we will analyze the 2D NDE and deduce predictions for the scaled spreading for two-dimensional lattices.

The NDE in terms of a time- and space-dependent energy density $\rho(\vec r,t)$ reads as:
\begin{equation} \label{eqn:NDE}
 \frac{\partial \rho}{\partial t} = \nabla \left(\rho^a \nabla \rho\right), 
 \qquad \mbox{with} \qquad \int \rho\, \rmd^2 \vec r = \energy\;.
\end{equation}
Here $\vec r$ is the two dimensional vector and $a$ is a nonlinearity index
 which later will be related to the exponents of the spreading.
The second equation represents the conservation of the total energy.
This density $\rho$ will be compared to the numerical results on the local energy density $w_{i,k}$ in the 2D lattices.
The NDE possesses a radially symmetric, self-similar solution:
\small
\begin{equation} \label{eqn:self_sim_solution_2d}
\rho = 
\begin{cases}
  (t-t_0)^{\frac{-1}{1+a}} \left(B - \frac{a}{4(a+1)} \frac{|\vec r|^2}{(t-t_0)^{\frac1{1+a}}} \right)^{\frac1a}& |\vec r|^2<R^2\\
    0 & |\vec r|^2>R^2
\end{cases}
\end{equation} \normalsize
with 
\begin{equation} \label{eqn:edge_propagation_2d}
  R^2 = 4B\frac{a+1}a \cdot (t-t_0)^\frac{1}{a+1} \quad\text{and}\quad B = \left(\frac{E}{4\pi}\right)^\frac{a}{a+1}.
\end{equation}
Here $R(t)$ denotes the radius of the excitation, hence the excitation area of spreading states should follow ${\dn\sim R^2}$.
Although this solution only solves~\eqref{eqn:NDE} for a source term $\rho(\vec r,t=0)=E\delta(\vec r)$ as initial condition, it was found that also quite arbitrary localized initial conditions  asymptotically converge towards this self-similar solution~\cite{Barenblatt_Zeldovich:72}.
Therefore, it can be viewed as describing the prototypical spreading behavior of the NDE.
Substituting $B$ in the expression for $R^2$, 
we find the scaling prediction of the NDE for the excitation area:
\begin{equation}  \label{eqn:scaled_spreading_2d}
 \frac{\dn}{E} \sim \left(\frac{t-t_0}{E}\right)^{\nu}\qquad \nu=\frac{1}{a+1}\;,
\end{equation} 
with $a$ being an unknown constant at this point.
However, for the homogeneous case $\kappa=\lambda$ it is possible to find an analytic prediction for $a$ as a function of the lattice nonlinearity $\kappa$ as will be presented in section~\ref{sec:nr}.
Moreover, for a regular quadratic local potential $\kappa=2$, $\omega_{i,k}=1$, all oscillators are in resonance and a resonant pertubation analysis gives an analytic prediction for $a$ in dependence of the coupling nonlinearity $\lambda$, also shown in the next section.

For the general, nonhomogeneous ($\kappa\neq\lambda$) disordered case $\omega_{i,k} \in [0,1]$ we still remain on the phenomenological level and check the scaling relation~\eqref{eqn:scaled_spreading_2d} by plotting the numerical data in appropriate coordinates. Generally, one may expect that $a$ is not a constant, but itself is a function of density $\nicefrac{E}{\dn}$.
In this case one would observe a deviation from the perfect power law~\eqref{eqn:scaled_spreading_2d} while the scaling coordinates may remain valid (such a situation was observed in several one-dimensional lattices~\cite{Mulansky-Ahnert-Pikovsky-11,Mulansky-Pikovsky-12}). 

From the self-similarity of the solution it immediately follows that the NDE predicts the structural entropy to be constant in the course of spreading:
\begin{equation}
 S_\text{str}(t) = S - S_2 \approx \text{const}\;.
\end{equation}
Deviations from this relation may indicate against validity of NDE.

\begin{figure}[t]
 \centering
  \includegraphics[width=0.48\textwidth]{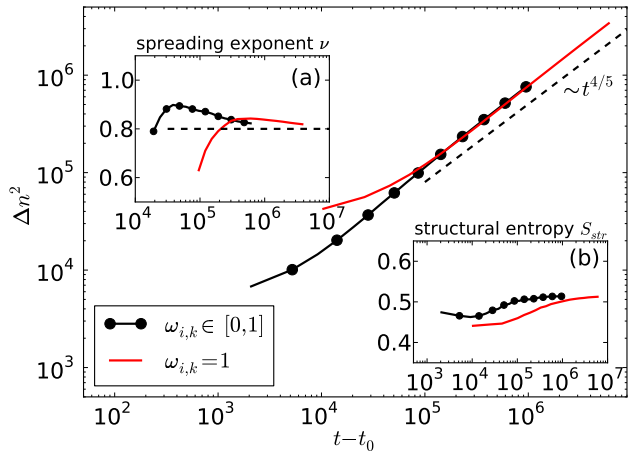} 
 \caption{(Color online) Second moment $\dn(t-t_0)$ for the homogeneous nonlinearities $\kappa=\lambda=4$ in a regular 2D ($\omega_{i,k} = 1$, red line) and a disordered ($\omega_{i,k}\in [0,1]$, black circle) lattice for $\beta=1$. $t_0$ was adjusted by hand to account for the long transient behavior before the power-law spreading is observed. The values for $t_0$ were about $t_0=10^4,\, 5\cdot10^4$ for the disordered and regular case.
The dashed lines show the expected behavior $\dn\sim (t-t_0)^{4/5}$. Note, that here no averaging over initial conditions was done, the graph shows the behavior of single trajectories. In inset (a) we plot the numerical spreading exponent $\nu$ obtained from finite differences, also together with the expectation from the NDE. Inset (b) shows the behavior of the structural entropy $S_\text{str}(t-t_0)$.
 }
 \label{fig:snol2d_4_4_dn2}
\end{figure}

\begin{figure}[t]
 \centering
  \includegraphics[width=0.48\textwidth]{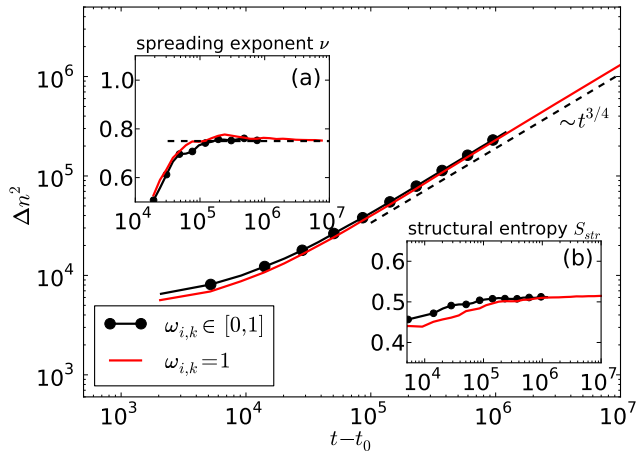}
 \caption{(Color online) Second moment $\dn(t-t_0)$ for the homogeneous case $\kappa=\lambda=6$ in a 2D lattice with disordered on-site potential ($\omega_{i,k} \in [0,1]$, black circles) and the regular case ($\omega_{i,k}=1$, red line). The transient time was adjusted as $t_0=10^4$. The dashed line shows the expected behavior $\dn\sim (t-t_0)^{3/4}$. In inset (a) we show the numerical spreading exponent $\nu$ which is the slope of the curves and should converge to $\nu=3/4$. Inset (b) shows the behavior of the structural entropy $S_\text{str}(t-t_0)$. Note that here no averaging over disorder realizations was done, the graphs show the behavior of single trajectories.
 }
 \label{fig:snhl2d_44_66_dn2_s}
\end{figure}

\mysection{Numerical Results}
\label{sec:nr}

\begin{figure}[t]
 \centering
  \includegraphics[width=0.48\textwidth,height=6cm]{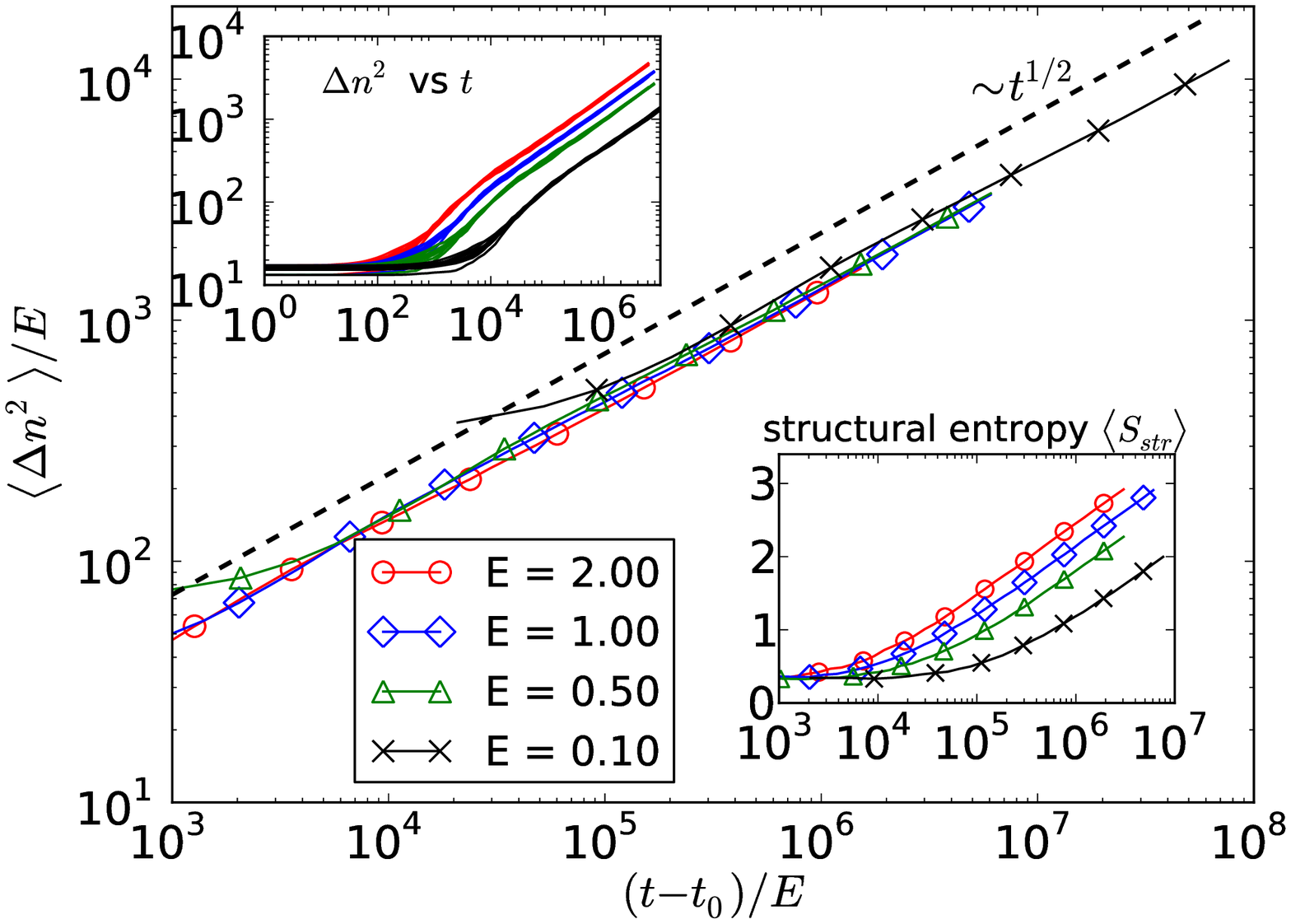} \\
  \includegraphics[width=0.48\textwidth,height=6cm]{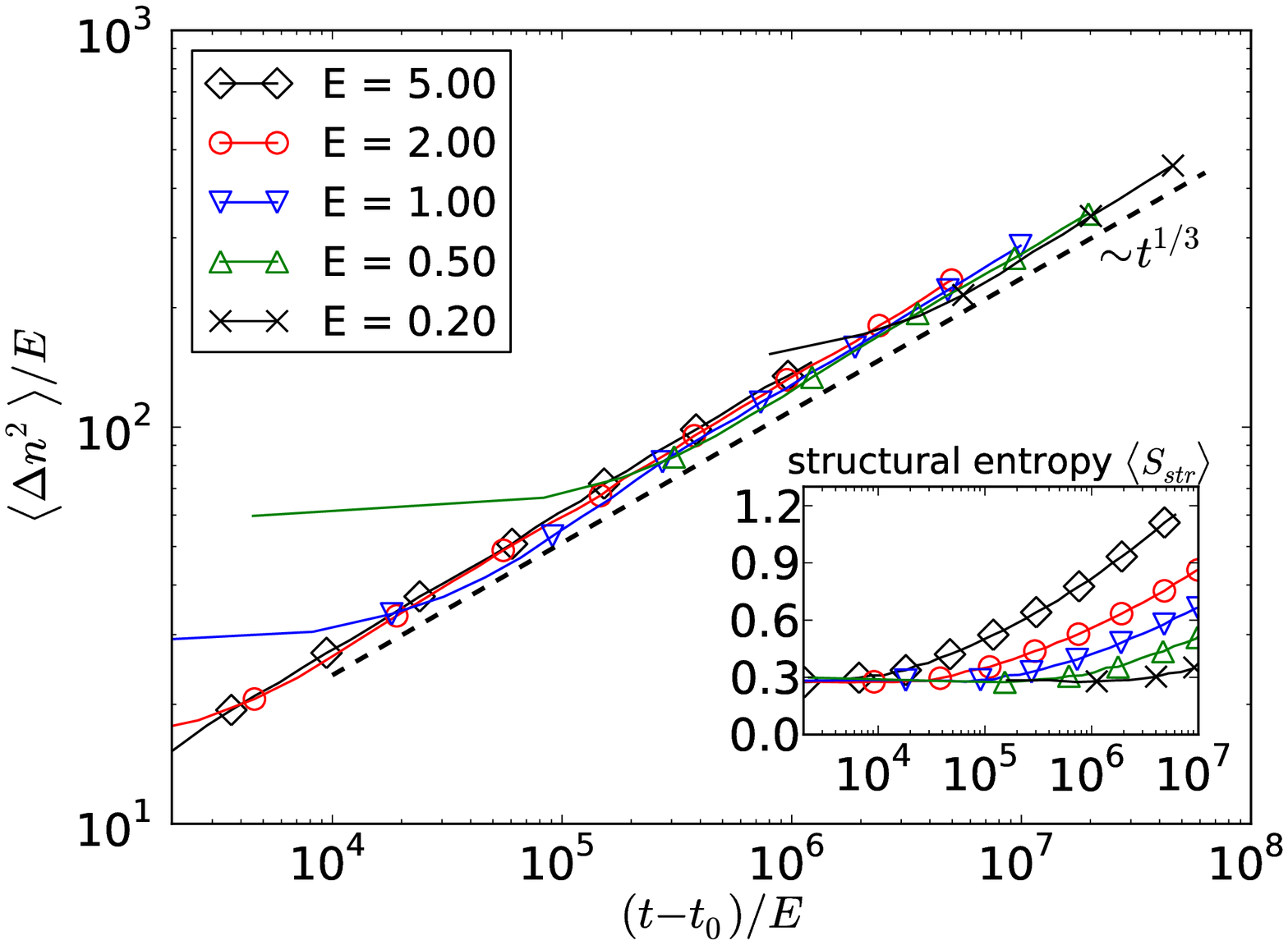}
 \caption{(Color online) Spreading results for the inhomogeneous case $\kappa=2$, $\lambda=4$ (upper graph) and $\lambda=6$ (lower graph) in a regular ($\omega_{i,k}=1$) lattice for different energies. The results are averaged over random initial conditions. We plot the scaling prediction $\langle\dn\rangle/E$ vs $t/E$. The upper inset for $\lambda=4$ shows the plain data, that is the second moment of $M=10$ trajectories for each energy (decreasing $E=2\dots0.1$) for lines from top to bottom in this plot ). The other insets show the behavior of the structural entropy $S_\text{str}(t)$.
 }
 \label{fig:snol2d_2_46_dn2}
\end{figure}

Here, we will check the analytical predictions from the NDE by numerical simulations of 2D nonlinear lattices of the form~\eqref{eqn:general_hamiltonian}.
For the numerical time evolution we used a 4-th order symplectic Runge-Kutta scheme~\cite{McLachlan-95} with step-size $\Delta t=0.1$ ($\Delta t=0.5$ in Fig.~\ref{fig:snol2d_4_4_dn2}), which conserved the energy with an accuracy of $\Delta E \lesssim 10^{-3}$.
The analytic scaling prediction~\eqref{eqn:scaled_spreading_2d} involves a parameter~$t_0$ that is introduced to account for transient behavior and is a priori unknown.
We have adjusted this parameter in the averaged results in a way to fully emphasize the transient behavior of the spreading.
The upper inset in Fig.~\ref{fig:snol2d_2_46_dn2} exemplarily shows the plain, unscaled results where no value $t_0$ is adjusted.
Note, that the values of $t_0$ are several orders of magnitude smaller than the maximum integration times and thus do not influence the asymptotic behavior at all.
This parameter is solely introduced to increase the visibility of the scaled plots by neglecting the transient behavior and emphasizing the asymptotic spreading.

\mysubsection{Homogeneous nonlinearity} 

At first, we chose the homogeneous case with $\kappa=\lambda=4,6$, where the energy can be set to unity and the only relevant parameter is the relative coupling strength $\beta$. 
First we note that from~\eqref{eqn:edge_propagation_2d} we find an energy-time relation, namely $t-t_0\sim \energy^{-a}$, that is imposed by the scaling property of the NDE~\cite{Mulansky-Ahnert-Pikovsky-11}.
Comparing this with the energy-time scaling in the homogeneous ($\kappa=\lambda$) lattice~\eqref{eqn:energy_time_scaling}, we get an exact result for the nonlinear parameter of the NDE in this case: $a=\frac{\kappa-2}{2\kappa}$.
Hence for $\kappa=\lambda$ the NDE gives an exact spreading prediction, namely:
\begin{equation}
 \dn \sim (t-t_0)^{2\kappa/(3\kappa-2)}.
 \label{eqn:hom}
\end{equation}

We compare this with numerical results in Figs.~\ref{fig:snol2d_4_4_dn2},\ref{fig:snhl2d_44_66_dn2_s}.
We used a random initial excitation of 5x5 sites, where $q_{i,k}=0$ and $p_{i,k}$ was chosen uniformly random iid., scaled in such a way that the total energy gave $E=1$.
We show results for $\beta=1$, additionally obtained results for $\beta = 0.1,2$ with similar outcome are ommited here.
For these simulations we show the results of a single, exemplary trajectory -- hence no averaging over random initial conditions is performed. Contrary to the one-dimensional case, here the fluctuations of the propagation velocity are very small, because of the effective averaging over the large border ($\sim 10^3$ sites) between excited and non-excited regions. 
Calculations for different random initial conditions with shorter integration times showed no significant difference.

In both cases we find a nice convergence of the spreading towards the predicted subdiffusive law $\dn \sim t^\nu$ with $\nu=2\kappa/(3\kappa-2)$, both for disordered and regular lattices.
Insets show the convergence of the spreading exponent obtained from finite differences of the numerical results on $\dn$.
For $\kappa=\lambda=4$, we have not quite reached the asymptotic regime in our simulation.
The results, however, seem to indicate a convergence towards the prediction $\nu=4/5$.
The integration for the regular lattices was chosen longer because of the longer transient observed there.
Another inset shows the structural entropy which converges to a constant value, in consistency with the NDE.

Overall, we believe that the good agreement of the numerical results with the predictions of the NDE in the homogeneous case $\kappa=\lambda$ is a convincing evidence that the NDE is the correct framework to describe spreading in nonlinear Hamiltonian lattices also in two dimensions.
It is quite remarkable that in this case the presence of disorder does not influence the asymptotic behavior at all, as seen from the perfect overlap of the curves in Figs.~\ref{fig:snol2d_4_4_dn2} and \ref{fig:snhl2d_44_66_dn2_s}.
This shows that disorder, despite some believe, is not essential for chaotic diffusion. However, such a result can only be found in two-dimensional systems, because in 1D disorder is required to block linear or nonlinear waves that would destroy the observation of diffusion.
We again note that in this case the NDE gives an exact analytic prediction on the spreading exponent $\nu$ which is verified numerically as the asymptotic behavior.

\mysubsection{Nonhomogeneous nonlinearity, regular lattice} 

We now turn to the more general situation of $\kappa \neq \lambda$ focusing on the case of the linear local oscillators $\kappa=2$ and nonlinear coupling $\lambda=4,6$.
We start with the regular lattice were $\omega_{i,k}=1$, hence all oscillators are in resonance.
In this case, again an analytic prediction for the NDE parameter $a$ can be obtained based on a time-scale analysis of the resonant dynamics.
The main idea is that the spreading can be considered as a sequence of excitation events: given a number of already excited oscillators, a new oscillator at the border will be subject to a resonant forcing induced by its already excited neighbor, and its energy will grow due to this forcing.

Let us consider an initially non-excited site, a neighbor of which demonstrates oscillations with amplitude $\epsilon$ and frequency $\Omega=1+a\epsilon^{\lambda-2}$ (this shift of frequency follows from the nonlinear coupling term). In the Hamiltonian for the initially non-excited site these oscillations appear as a driving force:
\begin{equation}
H_1=\frac{p^2+q^2}{2}+\frac{|q-\epsilon\sin\Omega t|^\lambda}{\lambda}\;.
\label{eqn:h1}
\end{equation}
To describe the resonant excitation, we transform to the action-angle 
variables $\theta$, $I$ with ${p=-\sqrt{2I}\cos(\theta-\Omega t)}$, ${q=\sqrt{2I}\sin(\theta-\Omega t)}$
to obtain:
\begin{equation}
H_1=-a\epsilon^{\lambda-2}I+\frac{|\sqrt{2I}\sin(\theta-\Omega t)-\epsilon\sin\Omega t|^\lambda}{\lambda}\;.
\label{eqn:h2}
\end{equation}
Averaging over the phase of fast oscillations $\Omega t$ we obtain a resonance averaged Hamiltonian:
\begin{equation}
\langle H_1\rangle=-a\epsilon^{\lambda-2}I+\epsilon^\lambda
F(\sqrt{2I\epsilon^{-2}}\cos\theta,\sqrt{2I\epsilon^{-2}}\sin\theta)
\label{eqn:h3}
\end{equation}
where $F(x,y)=\frac{1}{\lambda 2\pi}\int_0^{2\pi}d\xi\;|x\cos\xi-y\sin\xi-\sin\xi|^\lambda$.
The canonical equations of motion for $I,\theta$ can be reduced to a fully dimensionless form by rescaling $I\to \epsilon^2 I$, $t\to \epsilon^{\lambda-2}t$, dropping out any $\epsilon$--dependence. Thus, these equations describe growth of energy in the driven nonlinear oscillator to the level $ \sim \epsilon^2$ during the time $T\sim\epsilon^{2-\lambda}$. 

Let us now consider how this characteristic time scales with the total energy~$E$ in the lattice. If we assume that the form of the distribution of the energy over the excited sites remains the same for all energies, 
then $\epsilon^2\sim E$ and we obtain that the characteristic time for the excitation of an initially non-excited site scales with~$E$ as:
\begin{equation}
t\sim E^{\frac{2-\lambda}{2}}.
\label{eqn:h4}
\end{equation}
On the other hand, for the NDE~(\ref{eqn:NDE}) the relation $t\sim E^{-a}$ holds.
Thus, we obtain:
\begin{equation}
a=\frac{\lambda-2}{2} \quad \text{hence}\quad \Delta n^2 \sim t^{2/\lambda}.
\label{eqn:h5}
\end{equation}
We stress here again that this analysis is based on resonant excitation and definitely not possible neither for strongly nonlinear local potentials $\kappa>2$ nor for disordered lattices.

We check this prediction numerically for $\lambda=4,6$.
We started from $M=10$ ($M=24$ for $\lambda=6$) different random intial excitations of $5\times 5$ sites for each energy ($q_{i,k}=0$ as above), simulated the trajectories up a time $t_\text{end}=10^7$, and calculated the averages of $\dn$ and $S_\text{str}$.
Fig.~\ref{fig:snol2d_2_46_dn2} shows the results for the regular lattice ($\omega_{i,k}=1$) for $\kappa=2$ , $\lambda=4$ (upper graph) and $\lambda=6$ (lower graph) for different energies.
The inset in the upper graph shows the behavior of the second moment for all individual trajectories, i.e.~$M=10$ lines for each energy.
The asymptotically very small fluctuations for different initial conditions indicate that it is almost unneccessary to average at all for the regular case.
The NDE predicts the energy scaling of the spreading to be $\dn/E \sim ((t-t_0)/E)^{1/(a+1)}$, hence we adjust $t_0$ to account for the transient and plot $\dn/E$ vs $(t-t_0)/E$.
The values of $t_0$ were around $t_0\approx10^3$ for large energies and up to $t_0=5\cdot10^4$ for small energies, adjusted visually to emphasize the power-law behavior.
Indeed, this scaling gives an almost perfect collapse of data for different energies in both cases $\lambda=4,6$.
We also find perfect correspondance of the resonance prediction $a=(\lambda-2)/2$, hence $\Delta n^2 \sim t^\nu$ with $\nu = 2/\lambda$ from above~\eqref{eqn:h5}.

The structural entropy, however, shows a behavior in contradiction to the self-similar solution as seen in the insets of Fig.~\ref{fig:snol2d_2_46_dn2}.
We see a clear logarithmic increase of $S_\text{str}$ for all energies during the spreading.
This is a clear indication that the spreading state is not truly self-similar in time.
\new{An increase of the structural entropy means that the peaks in the energy distribution become statistically more pronounced.
It might thus be that some peaks in the distribution decay much slower than the average density, a ``breather-like'' behavior.}
As our measure of the excitation area $\dn$ is mainly governed by the boundary it is insensitive to the peak structure, which might explain why we still find the predicted scaling.
\new{However, in this work we can only report on those statistical properties of the change of the peak structure.
For a complete understanding a more detailed study is required that will be subject of future work.}

\mysubsection{Nonhomogeneous nonlinearity, disordered lattice} 

\begin{figure}[t]
 \centering
  \includegraphics[width=0.48\textwidth,height=6cm]{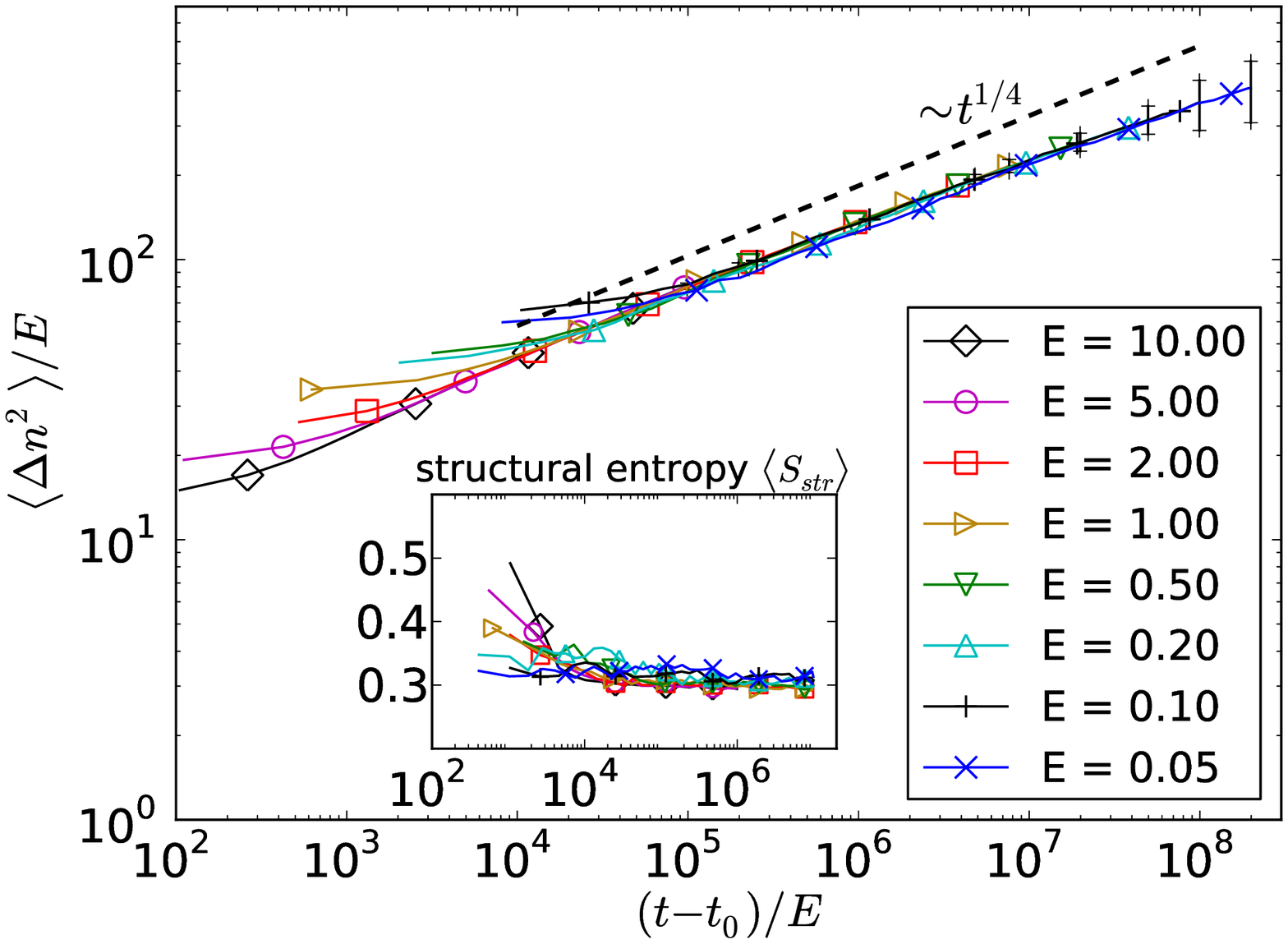} \\
  \includegraphics[width=0.48\textwidth,height=6cm]{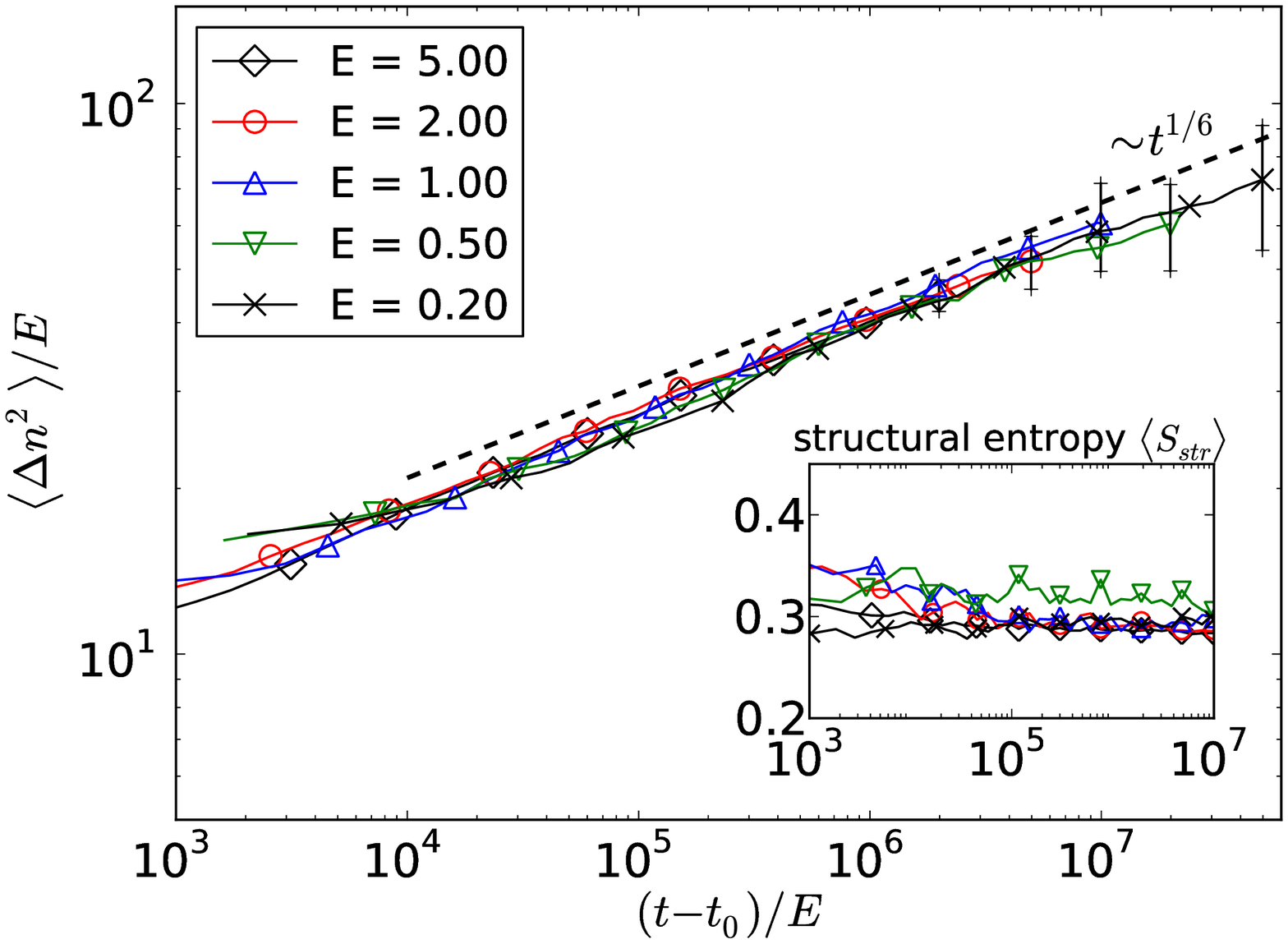}
 \caption{(Color online) Spreading results for the inhomogeneous case $\kappa=2$, $\lambda=4$ (upper graph) and $\lambda=6$ (lower graph) in a disordered ($\omega_{i,k} \in [0,1]$ iid.) lattice for different energies. The results are averaged over random potential realizations and the error bars indicate statistical errors. We plot the scaling prediction $\langle\dn\rangle/E$ vs $t/E$. The insets show the behavior of the structural entropy $\langle S_\text{str}\rangle(t)$.
 }
 \label{fig:snhl2d_46_dn2}
\end{figure}

Finally, we investigate the case of a disordered potential $\omega_{i,k} \in [0,1]$ also with $\kappa=2$, $\lambda=4$ and $\lambda=6$.
In this case the oscillators are (typically) out of resonance and the predictions above do not hold.
Generally, one expects a much slower spreading than for resonant oscillators.
A conclusive analytical estimation for this case is still lacking at this point.
Fig.~\ref{fig:snhl2d_46_dn2} shows the results for the disordered case for $\kappa=2$ , $\lambda=4$ (upper graph) and $\lambda=6$ (lower graph) for different energies.
Here, we average over at least 24 realizations of disorder and the total integration time was again $t_\text{end}=10^7$.
The disorder averaging here is essential, because of the very slow spreading which gives only about 100 excited sites at the end of simulation.
This is surely not enough for a reasonable self-averaging and hence the clarance of the results are improved by taken averages over disorder.
The errorbars in Fig.~\ref{fig:snol2d_2_46_dn2} represent the statistical error of the results and are plotted only for the final point at $t=10^7$ for each energy value.
The statistical error clearly increases for larger scaled time $t/E$ as for smaller energies the states occupy less lattice sites and hence one finds larger fluctuations.
The same argument explains why for $\lambda=6$ the statistical error is larger than for $\lambda=4$.
Again, we follow the prediction on the energy scaling of the NDE and plot $\dn/E$ vs $(t-t_0)/E$, where again $t_0$ was adjusted visually to values around $t_0\approx 5\cdot10^3$.
As above, we find an almost perfect collapse of data for different energies for the case $\lambda=4$, for $\lambda=6$ the collapse is not as clear but still convincing.
We observe a spreading exponent $\nu$ that is nearly constant $\nu_{4,d}\approx 1/4$ for $\lambda=4$ and $\nu_{6,r}\approx 1/6$ for $\lambda=6$ over the studied interval, with a slight decreasing of $\nu$ for long times/small energy densities.
Hence, the spreading process in the disordered case also obeys the energy scaling prediction of the NDE. Moreover, the saturation of the structual entropy is a sign of self-similarity of the spreading states.

The reason why, opposed to regular lattices, in the disordered case no change of the peak structure is observed is still to be understood.
One possible explanation might be that for a non-random potential one can apply an averaging over the fast oscillations of the oscillators which leads to a nonlinear Schr\"odinger lattice equation for the complex variable $\psi_{i,k} := q_{i,k} + \iota\, p_{i,k}$ \cite{Johansson-06}.
This introduces a new conserved quantity, $\norm = \sum |\psi|^2$, which supports the formation of breathers~\cite{Aubry-97}.
For a random potential, this averaging is not applicable, or at least leads to less exact results, as each oscillator has a random frequency and one can not apply a global averaging over these oscillations, hence breathers are less likely in this case.

\mysection{Conclusion}
In conclusion, we have extended the framework of the
phenomenological description
of energy spreading in strongly nonlinear lattices based on the self-similar solution of the Nonlinear Diffusion Equation to two dimensions.
This allowed for studying chaotic diffusion in the absence of disorder because in 2D lattice no non-linear waves seem to be present also for $\omega_{i,k}=1$.
The striking result is that one still observes chaotic diffusion very well described by the nonlinear diffusion equation, thus no disorder is required for this phenomenon.

For the homogeneous case of equal nonlinear powers $\kappa=\lambda$, it is possible to deduce an exact spreading law $\dn\sim t^\nu$ where the exponent can be calculated exactly from comparing the energy-time relations as $\nu=2\kappa/(3\kappa-2)$, similar to previous results in one dimension~\cite{Mulansky-Ahnert-Pikovsky-11}.
Numerical simulations of two-dimensional nonlinear Hamiltonian lattices with $\kappa=\lambda=4,6$ showed a good convergence of the spreading towards this analytic expectation.

For the non-homogeneous case, we checked the energy scaling prediction from the NDE for the two cases $\kappa=2$, $\lambda=4,6$  for disordered and regular lattices. 
We found that the spreading in such non-linear lattices does show the predicted scaling of the number of excited oscillators with energy and time. 
For the regular case there appears to be an additional mechanism that leads to deviations from the expected self-similarity of the field profile, as indicated by a growth of the structural entropy.
For disordered potentials, however, the numerical results for this entropy nicely follow the scaling prediction and we conclude that in this case the NDE again gives the correct description of the spreading behavior.
We have found that the effective index in the NDE for regular lattices ($a_{4,r}=1, a_{6,r}= 2$) is different from that in the disorderd case ($a_{4,d}\approx 3, a_{6,d}\approx 5$). Our explanation is that in the regular case, a resonant mechanism is mostly responsible for the spreading, as neighboring oscillators have close frequencies. Then from the scaling properties of the reduced Hamiltonian which describes the excitation of a new site, we derived a general relation between the power of NDE and the nonlinearity index in the lattice~(\ref{eqn:h5}) which yields correct values of $a_{4,r},a_{6,r}$. 
We note that this result is not based on the assumptions of  ``strong/weak chaos'' (cf.~\cite{Laptyeva-10,Flach-Krimer-Skokos-09}), but on the exact rescaling of the resonant Hamiltonian.
For other cases this resonance mechanism does not work, and a theoretical derivation of the relation between $a$ and $\lambda$ remains a challenge for future studies.

We note that for disordered lattice in one dimension, a deviation from the power law behavior was found with an energy dependent exponent $a(w)$, i.e. $\nu(w)$~\cite{Mulansky-Ahnert-Pikovsky-11}. The results in 2D seem not to verify this claim, although a careful look at the results for $\kappa=2$, $\lambda=4$ in Fig.~\ref{fig:snhl2d_46_dn2} do indicate a bending down of the curves.
There are two reasons why such a behavior is not observed here. Firstly, the excitation times used in~\cite{Mulansky-Ahnert-Pikovsky-11} are a much better quantity to identify such a density dependence due to the different averaging, as also explained there.
Moreover, in 2D each oscillator has not only one, but two or even three excited neighbors which might surpress the density dependence~\cite{Mulansky_phd:12}.
However, the exact role of the number of coupled neighbors is to be examined further in later works.


\acknowledgments
Most of the numerical results have been obtained at the CINECA sp6 supercomputer under the Project HPC-EUROPA2 (Project number 228398), with the support of the European Community - under the FP7 ``Research Infrastructure'' Programme.
M.~M.\ thanks the CNR Institute for Complex Systems in Florence for hospitality,  the IHP Paris for hospitality and financial support, and DFG for support under project PI 220/12-1. Fruitful discussions with D. Shepelyansky, S. Fishman, and S. Flach are cordially acknowledged.
Finally, we thank the referees for constructive feedback helping to improve the understandability of this text.


\end{document}